\documentclass[prb,superscriptaddress, preprintnumbers,twocolumn,showpacs]{revtex4}
\usepackage{amsmath}
\usepackage{amssymb}
\usepackage{amsthm}
\usepackage{amsfonts}
\usepackage{algorithmic}
\usepackage{enumerate}
\usepackage{latexsym}
\usepackage{graphicx}
\usepackage{bm}
\usepackage{subfigure}
\usepackage{ifthen}
\usepackage{sidecap}
\usepackage{txfonts}

\begin{document}

\title{The Coexistence and Decoupling of Bulk and Edge States in Disordered Two-dimensional Topological Insulators}
\author{Yan-Yang Zhang}
\affiliation{SKLSM, Institute of Semiconductors, Chinese Academy of
Sciences, P.O. Box 912, Beijing 100083, China}
\affiliation{Synergetic Innovation Center of Quantum Information and Quantum Physics, University of Science and Technology of China, Hefei, Anhui 230026, China}
\author{Man Shen}
\affiliation{SKLSM, Institute of Semiconductors, Chinese Academy of
Sciences, P.O. Box 912, Beijing 100083, China}
\affiliation{Department of Physics and Hebei Advanced Thin Film  Laboratory, Hebei Normal University, Hebei
050024, China}
\author{Xing-Tao An}
\affiliation{School of Sciences, Hebei University of
Science and Technology, Shijiazhuang, Hebei 050018, China}
\author{Qing-Feng Sun}
\affiliation{International Center for Quantum Materials and School of Physics, Peking University, Beijing 100871, China}
\author{Xin-Cheng Xie}
\affiliation{International Center for Quantum Materials and School of Physics, Peking University, Beijing 100871, China}
\author{Kai Chang}
\affiliation{SKLSM, Institute of Semiconductors, Chinese Academy of
Sciences, P.O. Box 912, Beijing 100083, China}
\affiliation{Synergetic Innovation Center of Quantum Information and Quantum Physics, University of Science and Technology of China, Hefei, Anhui 230026, China}
\author{Shu-Shen Li}
\affiliation{SKLSM, Institute of Semiconductors, Chinese Academy of
Sciences, P.O. Box 912, Beijing 100083, China}
\affiliation{Synergetic Innovation Center of Quantum Information and Quantum Physics, University of Science and Technology of China, Hefei, Anhui 230026, China}

\date{\today}

\begin{abstract}
We investigate the scattering and localization properties of edge
and bulk states in a disordered two-dimensional topological insulator when they
coexist at the same fermi energy. Due to edge-bulk backscattering
(which is not prohibited \emph{a priori} by topology or symmetry), Anderson
disorder makes the edge and bulk states localized indistinguishably.
Two methods are proposed to effectively decouple them and to restore
robust transport. The first kind of decouple is from long range disorder, since edge
and bulk states are well separated in $k$ space. The second one
is from an edge gating, owing to the edge nature of edge states in
real space. The latter can be used to electrically tune a system
between an Anderson insulator and a topologically robust conductor,
i.e., a realization of a topological transistor.
\end{abstract}

\maketitle

{\it Introduction.} What make topological insulators (TIs) unusual are the
boundary states carrying dissipationless currents. This novel
property is characterized by a topological invariant
(e.g., Chern number or $Z_2$ invariant) associated with the occupied
bulk bands. In the presence of boundaries (surfaces or edges), a
nontrivial topological invariant guarantees the existence
of gapless boundary states connecting the conduction and valance
bands\cite{Hasan2010,XLQiRMP,SQShen}. In the bulk gap, the absence of backscattering between
boundary states is protected by the topological order and some
symmetry. Topological states in
two-dimensional (2D) \cite{Konig2007,Roth2009,QKXue2013,RRDu2013}
and three-dimensional (3D)\cite{Hsieh2008,YXia2009} systems have
been experimentally observed recently.

However, the robust transport properties of boundary states are
only valid without the interference of bulk states. Unfortunately,
most of the 3DTIs found so far are actually metals, i.e., with fermi
energy in the bulk band coexisting with boundary states\cite{Hasan2010,XLQiRMP}.
This is one of the biggest experimental obstacles to realizing boundary transport in TIs. It
has been observed that the mix between boundary and bulk states
will lead to remarkable
backscattering\cite{Hasan2010,Hsieh2008,QKXue2010,SKim2011}. This
backscattering can destroy the perfect conducting of boundary
states. Therefore further understanding of boundary-bulk interplay
is necessary. In Refs.
\cite{Bergman2010,YTHsu2014}, Fano-like rearrangements of bulk and
boundary spectra arising from the mixing was discussed. Nevertheless,
the scattering and transport properties due to disorder is not
clear: how does such scattering happen? To what extent does it
affect the robustness of boundary states? Most importantly, how can
we avoid it?

In this paper, we systematically investigate the scattering
between edge and bulk states in 2DTIs, in the presence of
nonmagnetic impurities. We found that with the coexistence of edge
and bulk states at the same fermi energy: 1) Anderson disorder tends
to localize them indistinguishably; 2) Long range impurities can
effectively decouple them and restore perfectly conducting channels
(PCCs); 3) A local voltage gating at the edge is also enough to
effectively decouple them and leads to PCCs, therefore gives rise to
a convenient way to switch the system between a localized Anderson
insulator and a perfect conducting TI. This is a
concrete proposal of a topological field effect transistor (FET)
\cite{QKXue2011,LAWray2012}.

{\it Model and methods.} We adopt the Kane-Mele \cite{KaneMele2005} type model defined
on a honeycomb lattice
\begin{eqnarray}
H&=&t\sum_{\langle{ij}\rangle,\sigma}c_{i\sigma}^{\dag}c_{j\sigma}
+\lambda_{\nu}\sum_{i,\sigma}\xi_{i}c_{i\sigma}^{\dag}c_{i\sigma}\nonumber\\
&+&i\frac{\lambda_{\mathrm{SO}}}{3\sqrt{3}}\sum_{\langle\langle{ij}\rangle\rangle,\sigma\sigma^\prime}
\nu_{ij}c_{i\sigma}^{\dag}s^z_{\sigma\sigma^\prime}c_{j\sigma^\prime},
\label{eq1}
\end{eqnarray}
which has been used to describe the electronic states in silicene\cite{CCLiu2011B,Ezawa2012,Ezawa2012B}.
The first term describes the nearest-neighbor (NN) hopping , where $c_{i\sigma}^{\dag}$ creates an
electron at site $i$ with spin $\sigma$. The second term represents the staggered
potential with $\xi_{i}=\pm1$ for sublattice A (B). The third
term is the intrinsic spin-orbital coupling (SOC) between the near-nearest-neighbor (NNN) sites, where
$\bm{s}=(s_x,s_y,s_z)$ are the Pauli matrices for physical spins,
and $\nu_{ij}=(\bm{d}_i\times \bm{d}_j)_z/|\bm{d}_i\times
\bm{d}_j|=\pm 1$ with $\bm{d}_i$ and $\bm{d}_j$ the two NN bonds
connecting NNN sites $i$ and $j$.
Rashba SOC in realistic materials does not change the physics we will discuss,
therefore it is not included for simplicity.
Hereafter, we adopt
$t$ as the energy unit and lattice constant $a$ (NNN distance) as
the length unit.

\begin{figure}[tb]
\includegraphics*[width=0.4\textwidth,bb=0 50 1009 1080]{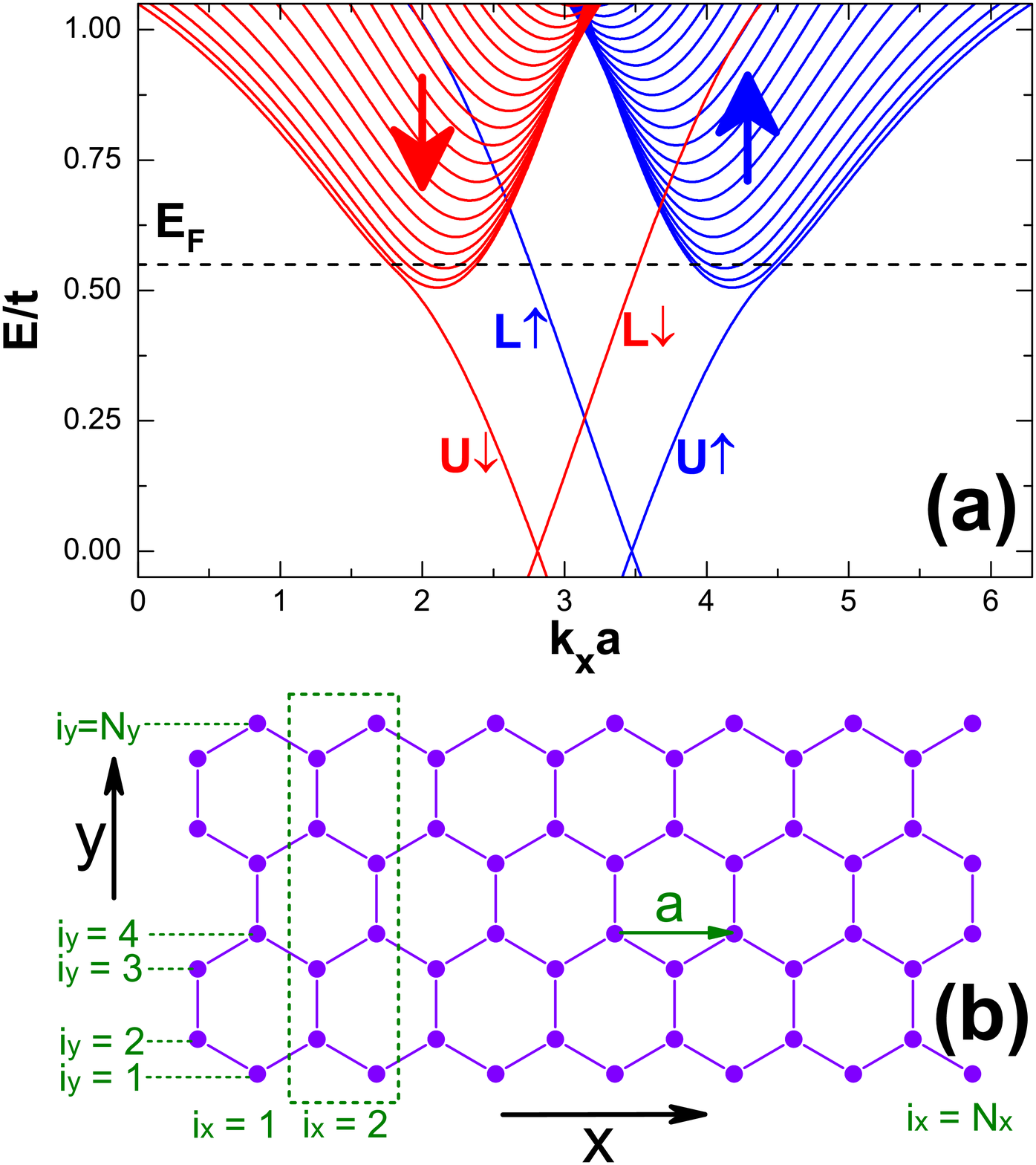}
\caption{(Color online) (a) Dispersion of a quasi-1D nanoribbon with
zigzag edges and finite width $N_y=80$. The model parameters are:
$\lambda_{\mathrm{SO}}=0.8t$ and $\lambda_{\mathrm{v}}=0.3t$. States with spin $\uparrow$ ($\downarrow$)
are plotted in blue (red). Edge states associated with lower (L) or
upper (U) edge are also marked. (b) Schematic of the zigzag edge nanoribbon with length $N_x$
and width $N_y$. The
unit cell is marked by the green dashed line. }
\label{FigDispersion}
\end{figure}

The electronic structure of the model (\ref{eq1}) has been
well studied\cite{Ezawa2012,Ezawa2012B}. The bulk bands are gapped
at Dirac points $K(K')=(\pm4\pi/3,0)$ with magnitude
$\Delta_G=2|\lambda_{\mathrm{SO}}-\lambda_{\mathrm{v}}|$. With
finite gap $\Delta_G$, the system is a topologically trivial
(nontrivial) insulator if
$\lambda_{\mathrm{SO}}<\lambda_{\mathrm{v}}$
($\lambda_{\mathrm{SO}}>\lambda_{\mathrm{v}}$).
In Fig. \ref{FigDispersion} (a), we plot the
typical band structure of a quasi-one dimensional (1D) ribbon in the
topological nontrivial phase. Within the bulk gap,
$|E|<|\lambda_{\mathrm{SO}}-\lambda_{\mathrm{v}}|$, there are four
gapless edge states, corresponding to lower (L) and upper (U) edges,
with spin up ($\uparrow$) and down ($\downarrow$), respectively. In
the presence of time reversal symmetry (TRS) and sufficiently ribbon
width, the backscattering between them is
prohibited\cite{BZhou2008}. This is a typical model of quantum spin
Hall effect.

The interests in the present work will be in the bulk bands. As shown in Fig. \ref{FigDispersion} (a), when
$E>\lambda_{\mathrm{SO}}-\lambda_{\mathrm{v}}=0.5t$, one pair of
edge states (U$\uparrow$ and U$\downarrow$) at the upper edge merge
into the bulk sates and lose their edge nature (hereafter they will be considered as bulk states),
while another pair of edge states (L$\uparrow$ and
L$\downarrow$) at the lower edge survives. These have also been checked by
calculating the wavefunction distributions in real space. In other
words, model parameters are chosen such that there are edge states
(well defined and distinguishable both in real space and in $k$ space) coexisting with the bulk
bands at the same fermi energy. In the following, we will
investigate the impurity scattering in this regime.

Non-magnetic impurities are expressed by adding a term to the
Hamiltonian as
\begin{equation}
H_{\mathrm{I}}=\sum_{i,\sigma}V_ic_{i\sigma}^{\dag}c_{i\sigma},\label{eqImpurity}
\end{equation}
where $V_i$ is the random onsite potential, and is independent
of spin to preserve TRS. Impurities induce scatterings between
states. We investigate this in a
standard geometry of quasi-1D ribbon infinitely extending in the $x$
direction, which is divided into three
parts: the left lead (source), the central region (sample),
and the right lead (drain). The leads are clean and semi-infinite,
with well-defined channels
indexed by $n$ at Fermi energy $E_F$. The width $N_y$ of
the ribbon is chosen to be sufficiently large to avoid finite-size effects\cite{BZhou2008}.
With impurities in the sample, the channel-resolved
transmission amplitudes $T_{mn}\equiv T_{m\leftarrow n}$ between
right-going channels, i.e., from channel $n$ in the left lead to
channel $m$ in the right lead, can be calculated by using the
numerical methods introduced in Refs. \cite{Ando1991,Khomyakov2005}.
The total transmission $\sum_{mn}T_{mn}$ is the conductance in units
of $e^2/h$ at zero temperature, with $m$ and $n$ running over all
right-going channels\cite{Imry1999}. The model parameters (except for
disorder) are chosen to be identical for the sample and leads to
reveal the intrinsic scattering behavior at this set of parameters.

In order to investigate the localization of edge states
specifically, we define the edge transmission as
\begin{eqnarray}
T_{\mathrm{edge}}&=&\sum_{m}\sum_{n\in \mathrm{edge}}T_{mn}
\end{eqnarray}
where $n$ only runs over \emph{edge} channels in the left lead, and $m$ runs
over all channels in the right lead. This $T_{\mathrm{edge}}$ reflects
the final fate of incident edge states after going through the sample.
The bulk transmission
$T_{\mathrm{bulk}}$ can be defined similarly. Due to
disorder, possible localization makes $T$'s badly distributed over
different disorder realizations. Thus it is essential to investigate
the geometric mean $T^{\mathrm{typ}}=\exp \langle \ln T \rangle$ to
give the ``typical value'' instead of the arithmetic mean
$T^{\mathrm{ave}}=\langle T \rangle$.\cite{KSlevin2001}

\begin{figure}[tb]
\includegraphics*[width=0.45\textwidth]{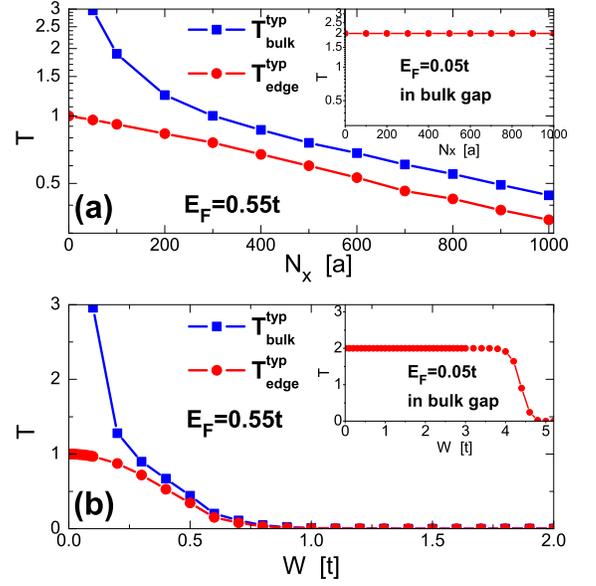}
\caption{(Color online) Typical values of bulk (blue square) and edge
(red circle) transmissions for Anderson disorder, with ribbon width $N_y=80$.
Main panels are for $E_F=0.55t$ (in band) while insets are for $E_F=0.05t$ (in bulk gap).
(a) Transmissions as functions of sample length $N_x$ with $W=0.5t$.
(b) Transmissions as functions of disorder strength $W$ with $N_x=1000a$.
Other model parameters are the same as in Fig. \ref{FigDispersion} (a). Each data
point is averaged over 1000 samples.  } \label{FigDisorder}
\end{figure}

{\it Anderson impurities.} In the
presence of bulk-edge coexistence, a natural conjecture is that
disorder will localize the bulk states and remain the edge states
left. We will show that the case is not so simple. It has also been
found that disorder can develop a topological phase from some definite Fermi energies in the bulk
band\cite{Li09,Jiang1,YXXing2011}. Starting
from a topologically nontrivial phase at clean limit, this just
originates from a particle-hole asymmetric band renormalization from
disorder\cite{SQShen,YYZhang2013}. Our context is distinct from those that at
clean limit, there exist well-defined
edge \emph{and} bulk states at the Fermi energy, as well as particle-hole symmetry.

Let's start from Anderson disorder, where the random onsite potential $V_i$
are independently and uniformly distributed between
$(-W/2,W/2)$. In the main panel of Fig. \ref{FigDisorder} (a), we show the bulk and
edge transmissions as functions of sample length $N_x$, at fermi
energy $E_F=0.55t$ in the conduction band, a little above the band edge $E_F=0.5t$.
Compared to the robust in-gap edge transmission shown in the inset,
the in-band transmissions are not robust and decay exponentially with increasing length.
With the $T-$axis (transmission) in a
logarithmic scale in Fig. \ref{FigDisorder} (a), the slope of the curve gives the inverse of
localization length $\lambda$. It is now clear that edge states and
bulk states localize with the same rate.

This non-robustness can also be seen from their dependence on disorder strength $W$, as shown in
Fig. \ref{FigDisorder} (b):
the in-band edge and bulk transmissions ($E_F=0.55t$, main panel) collapse to zero almost at the same time at $W\sim0.8t$, which are
much earlier than the robust in-gap ($E_F=0.05t$, inset) edge transmission at $W\sim5t$.
In this model, although
the backscattering from the edge state L$\downarrow$ (right-going) to the edge state L$\uparrow$ (left-going) is still prohibited by TRS,
the edge
state L$\downarrow$ still decays by leaking into the bulk and being
reflected to left going bulk channels. In other words, although
there exist topological edge states, their leakage and backscattering into the bulk
make the system into Anderson insulator, as normal disordered
systems in 2D.

\begin{figure}[tb]
\includegraphics*[width=0.45\textwidth]{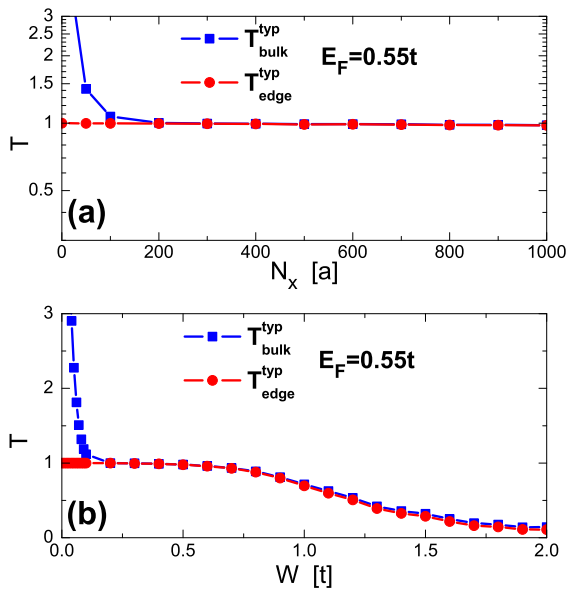}
\caption{(Color online) Same with main panels ($E_F=0.55t$) in Fig. \ref{FigDisorder} but for long
range impurities $\xi=1.5a$, $n_I=2.5\%$. (a) Transmissions as functions of sample length $N_x$ with $W=0.5t$.
(b) Transmissions as functions of disorder strength $W$ with $N_x=1000a$.  } \label{FigLongRange}
\end{figure}

{\it Long range impurities.} The above results show the disastrous consequence
of edge-bulk scattering in disordered 2DTI by destroying the robustness of edge states seriously.
We are now in a position to look for means to avoid this. As can be seen in Fig.
\ref{FigDispersion} (a), the bulk and edge states are well separated in
$k$ space when they coexist. To effectively decouple them, it is useful to
consider the long range disorder, which suppresses scattering
processes with large momentum transfer. In the presence of long
range disorder, the onsite potential $V_i$ is the sum of
contributions from $N_I$ impurities randomly centered at
$\{\mathbf{r}_m\}$ among $N$ sites
$V_{i}=\sum_{m=1}^{N_I}U_{m}\exp(-|\mathbf{r}_i-\mathbf{r}_m|^2/(2\xi))$,
where $U_{m}$ is randomly distributed within $(-W/2,W/2)$.
The above Anderson disorder corresponds to short range impurities with
correlation length $\xi=+0$ and density $n_i\equiv N_I/N=100\%$.
Impurities with correlation length $\xi>a$ can be said to be long
range, as widely investigated in graphene with substrate\cite{Ando1998,Rycerz07,Waka07}.

The transmissions for long range impurities ($\xi=1.5a$) are plotted in Fig. \ref{FigLongRange}.
In Fig. \ref{FigLongRange} (a), the edge transmission (red circle)
almost does not decay with length at all, restoring its robustness by effective decoupling
from bulk states. This robustness of the edge channel can also be seen from its dependence
on disorder strength $W$ in Fig. \ref{FigLongRange} (b).
On the other hand, the bulk transmission (blue
square) decays very fast before $N_x<200$, reflecting the Anderson
localization nature of the 2D bulk states. But after that this bulk
decay also ceases at another robust channel $e^2/h$. This residual
PCC can be attributed to the imbalance
between the number of left- and right-moving bulk
channels\cite{XRWang2004}, which is explained as follows.

The information of all active channels at
$E_F=0.55t$ are listed in Table \ref{Tab1}.
The long range nature of disorder effectively
decouples the edge channels (No. 8 for spin $\uparrow$ and No. 9
for spin $\downarrow$) from the bulk ones, which results in
an imbalance between opposite channels in the bulk: there is one more right
(left) going bulk channel for spin up
(down) component, and there is no scattering between states with opposite spins due to TRS.
Thus for a right
going setup here, the residual right going spin-up channel exhibits
itself by the appearance of a PCC. Notice
this PCC in the bulk is a collective effect, and it
does \emph{not} corresponds to any \emph{one} specific eigen-channel
listed in table \ref{Tab1}. Similar PCCs in other quasi-1D systems
with odd number of channels have been
discussed in Refs.\cite{Waka07,Takane2009}.

\begin{table}[htb]
\begin{tabular}{|c|c c c c c c c c|}
  \hline
  No. & 1 & 2 & 3 & 4 & 5 & 6 & 7 & 8 \\ \hline
  velocity & $-$ & $-$ & $-$ &$-$ & $+$ & $+$ & $+$ & $-$ \\
  spin & $\downarrow$ & $\downarrow$ & $\downarrow$ & $\downarrow$ & $\downarrow$ & $\downarrow$ & $\downarrow$ & $\uparrow$ \\
    &   &   &   &   &   &   &   & edge \\ \hline\hline
  No. & 9 & 10 & 11 & 12 & 13 & 14 & 15 & 16 \\ \hline
  velocity & $+$ & $-$ & $-$ & $-$ & $+$ & $+$ & $+$ & $+$ \\
  spin & $\downarrow$ & $\uparrow$ & $\uparrow$ & $\uparrow$ & $\uparrow$ & $\uparrow$ & $\uparrow$ & $\uparrow$ \\
    & edge &   &   &   &   &   &   &   \\
  \hline
\end{tabular}
\caption{Information for 16 conducting channels at $E_F=0.55t$ as marked in
Fig. \ref{FigDispersion} (a): the directions of their group velocities ($\pm$ for right (left) going)
and the orientations of their spins. The numbers are sorted by
ascending $k_x$.} \label{Tab1}
\end{table}

{\it Edge gating.} The decoupling of edge
and bulk states effectively recovers the robust conducting
behavior of topological states. The above decoupling from long
range disorder is based on the $k$ space consideration.
However, the correlation length $\xi$ of impurities depends strongly on the details of materials, and is hard to control. Now we
consider a real space proposal. For a 2DTI at the intrinsic Fermi energy $E_F=0.55t$ in Fig.
\ref{FigDispersion} (a), a simple way to discard the redundant bulk
states is to gate the whole 2D sample into the bulk gap, by a top gate covering the
whole sample sheet. However, we will show that,
an edge gating is also sufficient to create a
decoupling between edge and bulk states and therefore result in
PCCs. This is intuitive for a
generalization to 3DTIs, because a gating to a 3D sample can only
affect local carrier densities (therefore local chemical potentials) near the
interface to the gate, instead of those in the whole bulk.

\begin{figure}[htb]
\includegraphics*[width=0.45\textwidth]{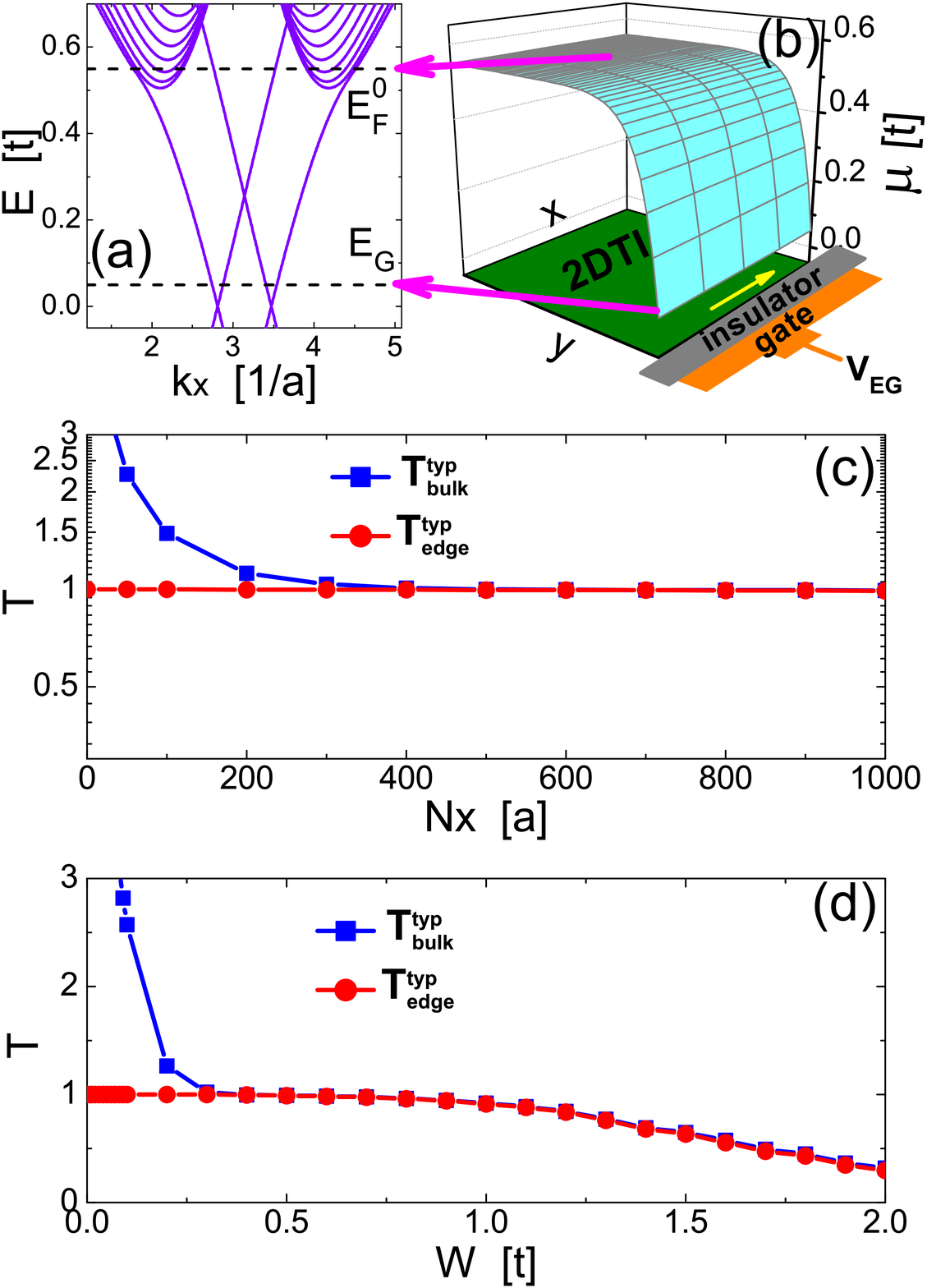}
\caption{(Color online) Edge gating. (a) Dispersion of the ribbon.
(b) Illustration of the metal edge gate (orange)/insulator (gray) structure attached to one edge of the 2DTI (green), and the configuration of chemical potential $\mu(x,y)$ it induces. The electronic transport is along $x$-direction. The yellow arrow represents the edge state.
(c) and (d) are all the same with main panels of Fig. \ref{FigDisorder} (a) and (b), but with an edge gating $E_G=0.05t$ and
$\xi_G=3a$.  }
\label{FigGate}
\end{figure}

As illustrated in
Fig. \ref{FigGate} (b), consider that a metal gate/insulator structure is attached to the
lower edge ($y=0$) of the 2DTI. This edge gate voltage $V_{EG}$ can only change the sample's
local chemical potential $\mu$ around this edge, say, in an exponential decaying form as
\begin{equation}
\mu(x,y)=E^0_F-(E^0_F-E_G) \exp (-y/\xi_G),
\end{equation}
where $E^0_F=\mu(y\gg\xi_G)$ is the intrinsic chemical potential (Fermi energy) in
the bulk, $E_G=\mu(y=0)$ is the chemical potential at the gated edge ($y=0$) determined by the edge gate $V_{EG}$, and $\xi_G$ is the
decay length. If $E_G$ can be tuned into the bulk gap, and $\xi_G$
is larger than the penetration depth of the edge state in transverse direction,
then there will be only edge state [the yellow arrow in Fig. \ref{FigGate} (b)] in this gated region ($0\leqslant y\lesssim
\xi_G $) near this edge, in spite of the presence of bulk states
outside it. Although at different chemical potentials, the gated
and ungated regions still belong to the same topological
phase, with no additional edge states at the
interface ($y\sim \xi_G$) between them. So far, edge and bulk states have
been well separated in real space. This decoupling should also restore the robustness of
transport even for Anderson disorder. This is confirmed from the transmissions shown in
Fig. \ref{FigGate} (c), where perfectly conducting edge and bulk
channels can be seen. The transmissions converge to $e^2/h$ rapidly
when $E_G$ goes into the bulk gap and when $\xi_G$ is just several
lattice constants long. We also show the transmissions at fixed length
$N_x=1000a$ when varying disorder strength $W$. As other proposals of edge tuning in 2DTI
\cite{YTZhang2011,XTAn2012,HCLi2012,HCLi2013}, we have made this
decoupling by taking advantage of the local nature of edge states.

Therefore, an edge gate can drive a disordered TI
with bulk-edge coexistence from Anderson insulator [Fig.
\ref{FigDisorder} (a)] to a perfectly conductor [Fig. \ref{FigGate}
(c)]. This transition can be experimentally tuned to make a
FET\cite{QKXue2011,LAWray2012}. The on\/off conductance ratio
$G_{\mathrm{on}}/G_{\mathrm{off}}\sim\exp[-N_x/\lambda_G]/\exp[-N_x/\lambda_0]]$,
where $\lambda_0$ and $\lambda_G$ are the localization length of
zero gating and edge gating respectively, as long as $N_x$ is within
the phase coherence length.
Moreover, the on state is very stable
due to its topological origin.
Since this tuning only
requires an electric control near the boundary, it can be
naturally generalized to 3D. Recently,
a 3DTI based FET with boundary gating has been experimentally
observed in \cite{HZhu2013}, and our present work offers a possible theoretical explanation.

{\it Conclusions and discussions.} All the above results are based on the
Hamiltonian (\ref{eq1}) for a topologically nontrivial phase $Z_2 =1$.
However, since this Hamiltonian does not contain Pauli matrices $s_x$ or $s_y$,
it can be decoupled into spin $\uparrow$ and spin $\downarrow$ sub-systems,
each of which is a Chern insulator with Chern number $\pm 1$\cite{XLQiRMP,Ezawa2012B}.
On the other hand, we have checked (but not shown here) that our results are also
valid in the presence of Rashba SOC\cite{Ezawa2012,Ezawa2012B}, which couples spin
$\uparrow$ and spin $\downarrow$. In one word,
our conclusions can be generalized to Chern insulators and general
$Z_2$ 2DTIs.

As a summary, we studied the edge
and bulk transmissions of 2DTI when they coexist at the same fermi
energy. Anderson disorder tends to localize edge and bulk states
with the same localization length, because of unprotected
backscattering between them. In order to decouple them and restore
the robust transport of edge states, we introduced long range
disorder and edge gating, based on momentum space and real space
considerations, respectively. By edge gating, one can electrically
tune the disordered system between an Anderson localized state and a
robust extended state, making it a topological FET.

This work was supported by NSFC (grant nos. 11374294, 11204294, 11274364) and 973 Program Project No. 2013CB933304.


\begin{thebibliography}{99}

\bibitem{Hasan2010} M. Z. Hasan and C. L. Kane, Rev. Mod. Phys. \textbf{82},
3045 (2010).

\bibitem{XLQiRMP} X.-L. Qi and S.-C. Zhang, Rev. Mod. Phys. \textbf{83},
1057 (2011).

\bibitem{SQShen} S.-Q. Shen, \emph{Topological Insulators: Dirac Equation in Condensed Matters}
,(Springer Berlin 2012).

\bibitem{Konig2007} M. K\"{o}nig, S. Wiedmann, C. Br\"{u}ne, A. Roth, H. Buhmann, L. W.
Molenkamp, X. L. Qi and S. C. Zhang, Science \textbf{318}, 766
(2007).

\bibitem{Roth2009} A. Roth,
C. Br\"{u}ne, H. Buhmann, L. W. Molenkamp, J. Maciejko, X.-L. Qi,
S.-C. Zhang, Science \textbf{325},  294 (2009).

\bibitem{QKXue2013} C.-Z. Chang,
J.-S. Zhang, X. Feng, J. Shen, Z.-C. Zhang, M.-H. Guo, K. Li, Y.-B.
Ou, P. Wei, L.-L. Wang, Z.-Q. Ji, Y. Feng, S.-H. Ji, X. Chen, J.-F.
Jia, X. Dai, Z. Fang, S.-C. Zhang, K. He, Y.-Y. Wang, L. Lu, X.-C.
Ma, Q.-K. Xue, Science \textbf{340}, 167 (2013).

\bibitem{RRDu2013} L.-J. Du, I. Knez, G. Sullivan, R.-R. Du, arXiv:1306.1925.

\bibitem{YXia2009} Y. Xia, D. Qian, D. Hsieh, L. Wray, A. Pal, H. Lin, A. Bansil,
D. Grauer, Y. S. Hor, R. J. Cava, and M. Z. Hasan, Nat.
Phys. 5, 398 (2009).

\bibitem{Hsieh2008} D. Hsieh, D. Qian, L. Wray, Y. Xia, Y. S. Hor, R. J. Cava, and M. Z.
Hasan, Nature \textbf{452}, 970 (2008).

\bibitem{QKXue2010} Y.-Y. Li, G. Wang , X.-G. Zhu, M.-H. Liu, C. Ye, X. Chen, Y.-Y.
Wang, K. He, L.-L. Wang, X.-C. Ma , H.-J. Zhang, X. Dai, Z. Fang,
X.-C. Xie, Y. Liu, X.-L. Qi, J.-F. Jia, S.-C. Zhang, and Q.-K. Xue,
Adv. Mater. \textbf{22}, 4002 (2010).

\bibitem{SKim2011} S. Kim, M. Ye, K. Kuroda, Y. Yamada, E. E. Krasovskii,
E. V. Chulkov, K. Miyamoto, M. Nakatake, T. Okuda, Y. Ueda, K.
Shimada, H. Namatame, M. Taniguchi, and A. Kimura, Phys. Rev. Lett.
\textbf{107}, 056803 (2011).

\bibitem{Bergman2010} D. L. Bergman and G. Refael, Phys. Rev. B \textbf{82}, 195417
(2010).

\bibitem{YTHsu2014} Y.-T. Hsu, M. Fischer, T. L. Hughes, K. Park, and E.-A. Kim,
arXiv:1401.3378.

\bibitem{QKXue2011} Q.-K. Xue, Nat. Nano. \textbf{6}, 197 (2011).

\bibitem{LAWray2012} L. Andrew Wray, Nat. Phys. \textbf{8}, 705 (2012).

\bibitem{KaneMele2005} C. L. Kane and E. J. Mele, Phys. Rev. Lett. \textbf{95}, 146802 (2005).

\bibitem{CCLiu2011B} C.-C. Liu, H. Jiang and Y. Yao, Phys. Rev. B \textbf{84},
195430 (2011).

\bibitem{Ezawa2012} M. Ezawa, New J. Phys. \textbf{14},
033003 (2012).

\bibitem{Ezawa2012B} M. Ezawa, Phys. Rev. Lett. \textbf{109},
055502 (2012).

\bibitem{BZhou2008} B. Zhou, H. Z. Lu, R. L. Chu, S. Q. Shen, and Q. Niu, Phys. Rev. Lett.
\textbf{101} 246807 (2008).

\bibitem{Ando1991} T. Ando, Phys. Rev. B \textbf{44}, 8017 (1991).

\bibitem{Khomyakov2005} P. A. Khomyakov, G. Brocks, V. Karpan, M. Zwierzycki, and P. J. Kelly,
Phys. Rev. B \textbf{72}, 035450 (2005).

\bibitem{Imry1999} Y. Imry and R. Landauer,
Rev. Mod. Phys. \textbf{71}, S306 (1999).

\bibitem{KSlevin2001} K. Slevin, P. Marko\v{s}, and T. Ohtsuki,
Phys. Rev. Lett. \textbf{86}, 3594 (2001).

\bibitem{Li09} J. Li, R.-L. Chu, J. K. Jain and S.-Q. Shen, Phys.
Rev. Lett. \textbf{102}, 136806 (2009).

\bibitem{Jiang1} H. Jiang, L. Wang, Q.-F. Sun, and X. C. Xie,
Phys. Rev. B \textbf{80}, 165316 (2009).

\bibitem{YXXing2011} Y.-X. Xing, L. Zhang and J. Wang, Phys. Rev.
B \textbf{84}, 035110 (2011).


\bibitem{Ando1998} T. Ando and Nakanishi, J. Phys. Soc. Jpn. \textbf{67}, 1704 (1998).

\bibitem{Rycerz07} A. Rycerz, J. Tworzyd{\l}o and C. W. J. Beenakker, Europhys. Lett. \textbf{79}, 57003 (2007).

\bibitem{Waka07} K. Wakabayashi, Y. Takane and M. Sigrist, Phys. Rev. Lett. \textbf{99}, 036601 (2007).

\bibitem{YYZhang2013} Y.-Y. Zhang and S.-Q.
Shen, Phys. Rev. B \textbf{88}, 195145 (2013).

\bibitem{XRWang2004} S. D. Wang, Z.-Z. Sun, G. Xiong, S. Yin, and X R Wang,
J. Phys. A: Math. Gen. \textbf{37} 1337 (2004)

\bibitem{Takane2009} Y. Takane, S. Iwasaki, Y. Yoshioka,
M. Yamamoto, and K. Wakabayashi, J. Phys. Soc. Jpn. \textbf{78} 034717 (2009).

\bibitem{YTZhang2011} Y.-T. Zhang, Q.-F. Sun and X. C. Xie, J. Appl. Phys. \textbf{109}, 123718 (2011).

\bibitem{XTAn2012} X.-T. An, Y.-Y. Zhang, J.-J. Liu, and S.-S. Li, New J. Phys. \textbf{14} 083039 (2012).

\bibitem{HCLi2012} H.-C. Li, L. Sheng and D. Y. Xing, Phys. Rev. Lett. \textbf{108} 196806 (2012).

\bibitem{HCLi2013} H.-C. Li, L. Sheng, R. Shen, L. B. Shao, B.-G. Wang, D. N. Sheng, and D.Y. Xing, Phys. Rev. Lett. \textbf{110} 266802 (2013).

\bibitem{HZhu2013} H. Zhu, C. A. Richter, E. Zhao, J. E. Bonevich, W. A. Kimes, H.-J. Jang,
H. Yuan, H. Li, A. Arab, O. Kirillov, J. E. Maslar, D. E. Ioannou, and Q. Li, Sci. Rep. \textbf{3}, 1757 (2013).

\end{thebibliography}
\end{document}